\documentclass{epl}
\usepackage{amsmath}
\usepackage{amsfonts}
\usepackage{amssymb}
\usepackage{graphicx}

\title{Temperature dependent transport of correlated disordered electrons: 
elastic vs. inelastic scattering}
\shorttitle{Temperature dependent transport of correlated etc.}
\author{M. C. O. Aguiar\inst{1,2} \and E.~Miranda\inst{2} \and V. 
Dobrosavljevi\'{c}\inst{1} \and E. Abrahams\inst{3} \and G.~Kotliar\inst{3}}
\shortauthor{M. C. O. Aguiar \etal}
\institute{
  \inst{1} Department of Physics and National High Magnetic Field Laboratory,
Florida State University - Tallahassee, FL 32306, USA\\
  \inst{2} Instituto de F\'{\i}sica Gleb Wataghin, Unicamp - C.P. 6165,
Campinas, SP 13083-970, Brazil\\
  \inst{3} Center for Materials Theory, Serin Physics Laboratory, Rutgers 
University - 136 Frelinghuysen Road, Piscataway, NJ 08854, USA
}

\begin{document}

\maketitle

\begin{abstract}
Temperature dependent transport of disordered electronic systems
is examined in the presence of strong correlations. In contrast to
what is assumed in Fermi liquid approaches, finite temperature
behavior in this regime proves largely dominated by inelastic
electron-electron scattering. This conclusion is valid in the
strong coupling limit, where the disorder, the correlations and
the Fermi energy are all comparable, as in many materials near the
metal-insulator transition.
\end{abstract}

Temperature dependence of transport is well understood in ordinary metals,
where it is dominated by electron-phonon scattering at room temperature.
Impurity scattering \cite{lee} becomes more important close to $T=0$ (where
phonons are frozen out), resulting in temperature-independent (residual)
resistivity. Weak temperature dependence in this regime reflects
multiple-scattering processes leading to so-called \textquotedblleft
quantum\textquotedblright\ corrections, including weak localization and
\textquotedblleft interaction\textquotedblright\ effects \cite{lee}.

Recent work \cite{zala} emphasized that these corrections reflect the
interference on Friedel oscillations produced by impurities embedded in an
electron gas. In this picture, temperature dependence emerges due to elastic
scattering off the screened impurity potential (which is temperature
dependent). The mechanism was argued \cite{zala} to apply equally well to both
the ballistic and the diffusive regime. In either case, however, these
processes are expected to dominate only if inelastic scattering plays a
sub-dominant role.

Renewed interest in these issues has resulted from recent
observations \cite{abrahams} of a surprisingly large drop of
resistivity at low temperatures in silicon MOSFET's . Because this
behavior begins to emerge already at temperatures comparable to
the Fermi energy ($\sim$10K), estimates show \cite{abrahams} that
it takes place in the ballistic regime. Accordingly, several
authors \cite{dassarma,dolgopolov,herbut,zala} have proposed that
this reflects temperature dependent screening of the random
potential. On the other hand, the phenomenon is believed
\cite{abrahams,mitglass} to occur in the strongly correlated
regime, where inelastic electron-electron scattering may be
equally important.

In this Letter, we address the importance of inelastic processes
as a competing mechanism to temperature-dependent elastic
scattering off the screened impurity potential. This is difficult
within Fermi liquid approaches \cite{lee, zala}, which implicitly
\textit{assume} the sub-dominance of inelastic processes. A
framework where this general question can be answered in a precise
and controlled fashion is provided by Dynamical Mean Field Theory
(DMFT) \cite{georges}, which describes both the elastic and the
inelastic processes on the same footing. Our results demonstrate
that: (i) In the regime of strong correlation, where the
interaction, disorder, and the Fermi energy are all comparable, there
is a surprisingly large drop of resistivity (up to a factor of
ten or more). (ii) Here, Fermi liquid coherence occurs only at
rather low temperatures, while strong inelastic electron-electron
scattering (leading to decoherence) sets in rapidly as the
temperature is raised \cite{decoherence}. In fact, inelastic
processes completely dominate the \emph{entire} temperature regime
where the large resistivity drop is found ($0.04\leq
T/E_{F}\leq0.3$).
\begin{figure}[ptbh]
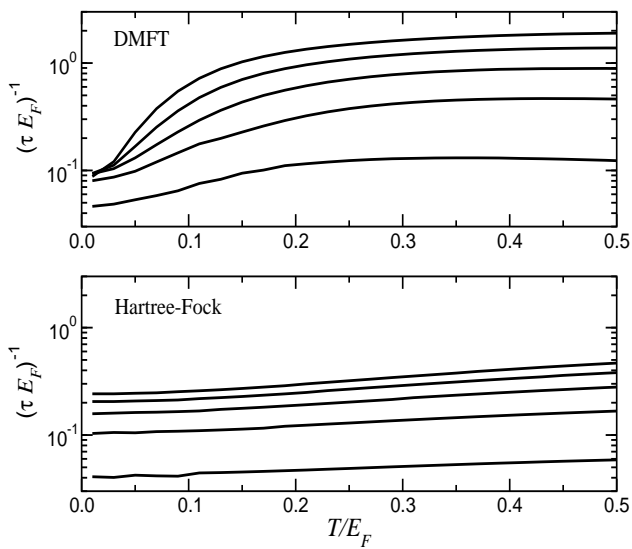

\onefigure[
trim=0.000000in 0.000000in 0.000000in -0.346384in,
height=2.9572in, width=3.2629in ]{fig1.eps}
\caption{Scattering rate $\tau^{-1}$($\sim$ resistivity) as a
function of temperature. Results are shown for the disorder
strength $W$ equal to the interaction $U$, as we reduce the Fermi
energy $E_{F}/U=$ 2.0, 1.0, 0.67, 0.5, 0.4 (bottom to top curves).
Note the large resistivity drop in the DMFT solution (top), but
much weaker
temperature dependence within the Hartree-Fock approach (bottom).}%
\label{f.1}%
\end{figure}

\textit{Finite temperature DMFT for disordered electrons. }We considered a
half-filled Hubbard model in the presence of random site energies, as given by
the Hamiltonian
\begin{equation}
H=-t\sum_{<ij>\sigma}c_{i\sigma}^{\dagger}c_{j\sigma}+\sum_{i\sigma}%
\epsilon_{i}n_{i\sigma}+U\sum_{i}n_{i\uparrow}n_{i\downarrow}. \label{e.1}
\end{equation}
Here $c_{i\sigma}^{\dagger}$ ($c_{i\sigma}$) creates (destroys) a
conduction electron with spin $\sigma$ on site $i$,
$n_{i\sigma}=c_{i\sigma}^{\dagger }c_{i\sigma}$ is the particle
number operator, $t$ is the hopping amplitude, and $U$ is the
on-site repulsion. The random site energies $\varepsilon_{i\text{}}$ 
are assumed to have a uniform distribution of width $W$. Within
DMFT for disordered electrons \cite{dmftdis}, a quasiparticle is
characterized by a local but site-dependent \cite{zimanyi}
self-energy function $\Sigma
_{i}(\omega)=\Sigma(\omega,\varepsilon_{i})$. To calculate these
self-energies, the problem is mapped onto an ensemble of Anderson
impurity problems \cite{dmftdis} embedded in a self-consistently
calculated conduction bath. In this approach, only quantitative
details of the solution depend on the details of the electronic
band structure; in the following we concentrate on a semi-circular
model density of states. To solve DMFT equations at finite
temperature, we mostly used the iterated perturbation theory (ITP)
method of Kajueter and Kotliar~\cite{kajueter}. However, we
carefully checked that all the qualitative features that we report
also appear when we solve DMFT equations using a Quantum Monte
Carlo impurity solver \cite{georges}.
\begin{figure}[ptbh]
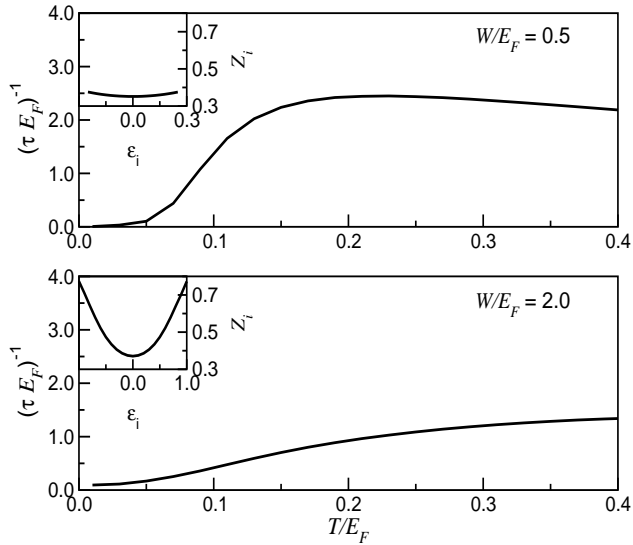

\onefigure[trim=0.000000in 0.000000in 0.000000in -0.330509in,
height=2.9572in,width=3.2629in]{fig2.eps}
\caption{Disorder-dependence of the scattering rate for $U=2E_{F}$, evaluated
for weak ($W=0.5E_{F}$; top) and strong ($W=2E_{F}$; bottom) disorder. Insets
show the distributions of local quasiparticle weights $Z_{i}$. }%
\label{f.2}%
\end{figure}

\textit{Temperature-dependent scattering rate.} Within DMFT
\cite{georges}, the temperature dependence of the resistivity
essentially follows that of the
total scattering rate, which takes the form $\tau^{-1}=-2\operatorname{Im}%
\Sigma_{av}(\omega=0)$, where the \textquotedblleft average\textquotedblright%
\ self energy \cite{dmftdis} corresponds to the disorder-averaged local
Green's function $\overline{G}(\omega)=\left\langle G_{i}(\omega)
\right\rangle _{\varepsilon_{i}}=G_{o}\left[  \omega-\Sigma_{av}%
(\omega)\right]  $, and $G_{o}(\omega)$ is the \textquotedblleft
bare\textquotedblright\ Green's function evaluated at $U=W=0.$ To
examine the effect of strong correlations on transport, we first
concentrate on the experimentally relevant regime where the
disorder and the correlations are comparable. We set $U=W$, and
examine the evolution of $\tau^{-1}(T)$ as the Fermi energy is
gradually reduced. Typical results of DMFT calculations are shown
in fig.~\ref{f.1} (top). We find that, as soon as the interaction $U$ is
comparable to electronic bandwidth $B$ (at half-filing
$B=2E_{F}$), the scattering rate displays a dramatic drop (of
order ten!) below temperatures $T\sim0.3E_{F}$, very similar to
the experiments \cite{abrahams}. We contrast this result to that
of standard weak-coupling approaches
\cite{dassarma,dolgopolov,herbut}, where the temperature
dependence is much weaker and occurs over a very broad temperature
range set simply by the bare Fermi scale. To make this comparison
more precise, we solve our DMFT equations by using the
Hartree-Fock (HF) approximation \cite{herbut} where $\Sigma
_{i}(\omega)=U\,n_{i}$; the results are shown in fig.~\ref{f.1} (bottom).
Very weak temperature dependence is found, and one has to go to
very high temperature ($T\gg E_{F}$) to get an appreciable rise in
resistivity. Note that, while giving much higher resistivity at
higher temperatures, the DMFT method also produces appreciably
lower resistivity at $T=0$, consistent with the phenomenon of
correlation-enhanced screening of the random potential
\cite{screening}.
\begin{figure}[ptbh]
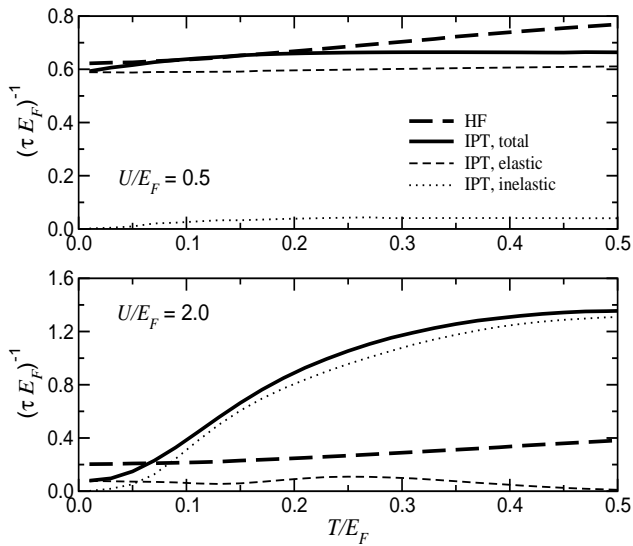

\onefigure[trim=0.000000in 0.000000in 0.000000in -0.346384in,
height=2.9572in,width=3.2629in]{fig3.eps}
\caption{DMFT results for $W=2E_{F}$ showing the total, elastic, and inelastic
scattering rates as functions of temperature are compared to predictions of
the Hartree-Fock (HF) approximation. Inelastic scattering dominates in the
strongly correlated limit.}%
\label{f.3}%
\end{figure}

\textit{Gradual decoherence due to disorder.} What sets the energy
scale for the resistivity drop? To answer this, we contrast
results obtained for $U=2E_{F}$ at weak and strong disorder, as
shown in fig.~\ref{f.2}. At weak disorder ($W=0.5E_{F}$), the behavior is
similar to the clean case \cite{georges}, where a dramatic
resistivity rise is found above a well defined
\textquotedblleft decoherence\textquotedblright\ temperature $T^{\ast}%
\approx0.1ZE_{F}$ \cite{dmftdis}, where the quasiparticles become
ill-defined (here $Z$ is the quasiparticle weight, see below).
This behavior is characteristic of many strongly correlated
systems such as heavy fermion compounds, but such an extremely large rise
is not seen in two-dimensional electron gases \cite{abrahams}. On
the other hand, our results for the moderately disordered
situation ($W=2E_{F}$) show a much more gradual resistivity rise,
as seen in fig.~\ref{f.2} and in the 2D experiments. To understand this
behavior, we note that in correlated disordered systems the
quasiparticle weight becomes a strongly site-dependent
\cite{zimanyi,dmftdis} quantity $Z_{i}$, which in the
DMFT limit is defined by%
\begin{equation}
Z_{i}=\left[  \left.  1-\frac{\partial}{\partial\omega}\mbox{Im}\Sigma
_{i}(\omega)\right\vert _{\omega\rightarrow0}\right]  ^{-1}. \label{e.2}
\end{equation}
The insets in fig.~\ref{f.2} show the distributions of
$Z_{i}=Z(\varepsilon_{i})$ for the two cases. For $U=2E_{F\text{
}}$and weak disorder, the $Z_{i}$-s are narrowly distributed
around $Z\approx0.36$ (corresponding to a mass enhancement
$m^{\ast}/m=Z^{-1}\approx2.8,$ and a decoherence temperature
$T^{\ast}/E_{F\text{ }}\approx0.04$), giving rise to a sharply
defined decoherence scale. In contrast, for stronger disorder the
$Z_{i}$-s are distributed over a broad interval $0.37<Z_{i}<0.8,$
corresponding to a broad distribution of local decoherence scales
$T_{i}^{\ast}\approx0.1Z_{i}E_{F}$. As the temperature is raised
in the presence of strong disorder, more and more sites gradually
become incoherent and act as strong scattering centers. If the
distribution $P(T_{i}^{\ast})$ is broad, then at intermediate
temperatures $T_{\min}^{\ast}<T<T_{\max}^{\ast}$ we expect the
density of such scattering centers to grow linearly with
temperature, resulting in a roughly linear resistivity in this
range. This behavior is indeed observed in our calculation of the
full scattering rate, which for strong disorder shows a roughly
linear dependence over an appreciable range. Interestingly, the
overall shape of this temperature dependence looks remarkably
similar to the experimental data on silicon MOSFET's
\cite{abrahams}.
\begin{figure}[ptbh]
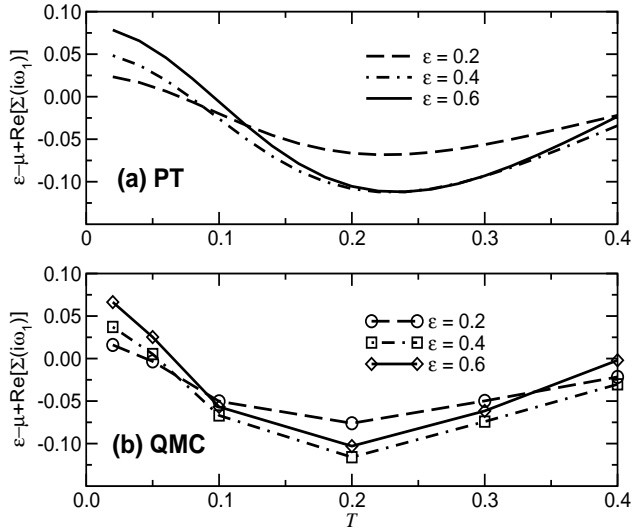

\onefigure[trim=0.000000in 0.000000in 0.000000in -0.523615in,
height=2.9572in,width=3.2629in]{fig4.eps}
\caption{Non-monotonic temperature dependence of renormalized (screened) site
energies $v_{i}(T)$ (see text) in the strongly correlated regime. Results
shown correspond to a simple Anderson-impurity model with $U=3$ in a
featureless host ($E_{F}=1$), using the IPT (top) and QMC (bottom) as the
impurity solver.}%
\label{f.4}%
\end{figure}

\textit{Elastic or inelastic scattering?} The total scattering rate $\tau
^{-1}$ which we have calculated describes the contribution of both the elastic
and the inelastic scattering. However, to better understand which of these two
processes dominates, we will separately estimate each of these contributions,
as follows. Both are completely determined by the zero frequency limit of the
local self-energy function, viz. $\Sigma_{i}(T)=\lim_{\omega\longrightarrow
0}\Sigma(\omega,\varepsilon_{i})$. Its real part determines the renormalized
(screened) random potential \cite{screening} $v_{i}(T)=\varepsilon
_{i}+\operatorname{Re}\Sigma_{i}(T)$, while the imaginary part describes the
local inelastic scattering rate $\tau_{inelastic}^{-1}(i)=-2\operatorname{Im}%
\Sigma_{i}(T)$ (which is nonzero only at $T>0)$. Using our self-consistent
procedure, we explicitly calculate both $\operatorname{Re}\Sigma_{i}(T)$ and
$\operatorname{Im}\Sigma_{i}(T)$ at a given temperature $T,$ as functions of
the local site energy $\varepsilon_{i}$. Once these quantities are known, we
can estimate the elastic (inelastic) contribution to the total scattering rate
by simply dropping the imaginary (real) part of $\varepsilon_{i}+\Sigma
_{i}(T)$, before computing $\overline{G}(\omega)$ from which $\tau
^{-1}=-2\operatorname{Im}\Sigma_{av}(\omega=0)$ is calculated.

In this way we have (for $W=2E_{F})$ computed the total, the elastic, and the
inelastic scattering rates as functions of temperature, as shown in fig.~\ref{f.3}. It
is also instructive to compare our DMFT results to those obtained for the same
model using the HF approximation. This weak-coupling approach is similar to
those used by most other theories \cite{dassarma,dolgopolov,herbut,zala},
which largely ignore the inelastic scattering. We find that in the weakly
interacting limit ($U=0.5E_{F}$; top panel) the elastic scattering dominates,
and good agreement is found between DMFT and HF predictions. However, when
strong correlations are present ($U=2E_{F}$; bottom panel), the inelastic
scattering proves much larger then the elastic component for all except the
lowest temperatures (elastic and inelastic contributions become comparable
around $T/E_{F\text{ }}\sim0.04$). These results demonstrate that
\textit{inelastic scattering dominates over the entire temperature range where
the large resistivity drop is found,} in striking contrast to weak-coupling
predictions \cite{dassarma,dolgopolov,herbut,zala}. In this regime, the
elastic scattering component has an extremely weak and even non monotonic
temperature dependence, and clearly has very little physical content if
considered in absence of inelastic processes.

To clarify this issue, we have explicitly computed the temperature
dependence of the renormalized site energies $v_{i}(T)$. In the
weakly interacting limit, these quantities are found to have a
modest and monotonic temperature dependence in agreement with HF
predictions. However, in the regime of strong correlations, we
find surprising non-monotonic temperature dependence where for
some values of $\varepsilon_{i}$, and at intermediate
temperatures, \textit{negative screening} is found ($v_{i}(T)<0$
for $\varepsilon_{i}>0$). We have examined this puzzling behavior
in great detail, and have found that this is a very general
feature of strongly correlated systems, which depends only weakly
on the specific self-consistency condition used.

To illustrate this, in fig.~\ref{f.4} we present results of such a calculation for a
simple Anderson impurity model with bare site energy $\varepsilon_{i}$ and
on-site repulsion $U$, embedded in a featureless (semicircular) electron bath.
To demonstrate that this behavior is not an artifact of our IPT impurity
solver, we present results of both IPT and numerically exact QMC calculations
for the same model, which produce almost identical results.

We emphasize that the energy scale associated with the $v_{i}(T)$-s is very
small in the intermediate temperature range where negative screening emerges.
In this regime, the scattering is completely dominated by inelastic processes,
so this puzzling behavior has by itself very little physical consequence.
Nevertheless, these results clearly indicate that theories which ignore
inelastic scattering are very likely to produce unreliable and even physically
incorrect results if used in the incoherent regime where Fermi liquid theory
cannot be applied. 
\begin{figure}[ptbh]
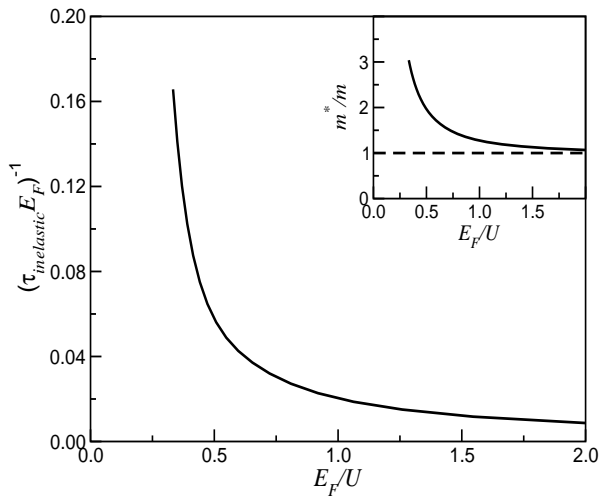

\onefigure[trim=0.000000in 0.000000in 0.000000in -0.621020in,
height=2.8in,width=3.1in]{fig5.eps}
\caption{Inelastic scattering rate for $W=U$ and $T/E_{F\text{ }}=0.05,$ as a
function of $E_{F}/U$. The inset shows the (average) effective mass
enhancement $m^{\ast}/m=\left\langle Z_{i}\right\rangle ^{-1}$ in the same
range of parameters. Large enhancement of dephasing is found in the same range
where the mass enhancement is large, similarly as in experiments.}%
\label{f.5}%
\end{figure}

\textit{Enhanced dephasing in the strongly correlated regime. }In
fig.~\ref{f.3}, we showed the inelastic scattering rate as a function of
temperature. We have also computed it as a function of density
(i.e. $E_{F}/U$) at $T/E_{F\text{ }}=0.05$ and $W=U$. As shown in
fig.~\ref{f.5}, the inelastic scattering rate becomes appreciable in the
same range where the effective mass ($m^{\ast}/m\sim\left\langle
Z\right\rangle ^{-1}$; see inset) is enhanced, and we enter the
regime of strong correlations. This prediction awaits experimental
tests on sufficiently homogeneous samples.

In summary, we have presented quantitatively reliable model
calculations for correlated disordered electrons in the strong
coupling limit where the disorder strength, the interactions, and
the Fermi energy are all comparable. Our results demonstrate that
inelastic electron-electron scattering dominates the regimes
relevant to many puzzling experimental situations. We expect large
resistivity drops, similar to what we find, to also occur in 3D
situations when correlations are sufficiently large, and diagonal
disorder not too strong \cite{sip}. In fact, in many 3D weakly
disordered heavy-fermion compounds, even larger resistivity drops
are observed below a coherence temperature. Our DMFT approach,
while being able to address the nontrivial interplay of disorder
and strong correlations is nevertheless too simple to include
localization effects that are important closer to the
metal-insulator transition. These can be incorporated in
our framework using recently developed extensions \cite{motand} of
DMFT, but this problem remains a fascinating direction for future
work.

\acknowledgments
We thank M.J. Rozenberg for providing us with the QMC code for the Hubbard
model, and A. Georges, S. Das Sarma, D. Popovi\'{c} for useful discussions.
This work was supported by FAPESP 99/00895-9 (MCOA), 01/00719-8 and CNPq
301222/97-5 (EM), and NSF grants DMR-9974311 and DMR-0234215 (VD), DMR-9976665
(EA), and DMR-0096462 (GK). VD and GK also thank KITP at UCSB (NSF grant
PHY99-07949) where part of this work was carried out.


\begin{thebibliography}{99}
\bibitem{decoherence} This feature is common to all theories of metals with
strong correlation.
\bibitem{sip} The well-studied MIT in Si:P does not meet the mentioned
criteria since there one is not close to the clean Mott
transition. In addition, Si:P has rather strong off-diagonal
disorder that is not moderated by the screening effects which are
important in our calculation.
\end{thebibliography}
\end{document}